\documentclass[12pt,a4paper]{article}
\usepackage{amsmath,amssymb,type1cm,booktabs,bm,braket,mymacros}
\allowdisplaybreaks[1]
\numberwithin{equation}{section}
\begin{document}
\begin{titlepage}

 \begin{flushright}
 \end{flushright}

 \vspace{1ex}

 \begin{center}

  {\LARGE\bf Flavor structure of $E_6$ GUT models}

  \vspace{3ex}

  {\large Hidetoshi Kawase
  \footnote{e-mail: hkawase@eken.phys.nagoya-u.ac.jp},
  Nobuhiro Maekawa
  \footnote{e-mail: maekawa@eken.phys.nagoya-u.ac.jp}}

  \vspace{4ex}
  {\it Department of Physics, Nagoya University, Nagoya 464-8602, Japan}
  \\
  \vspace{6ex}

 \end{center}

 \begin{abstract}
  It has been pointed out that in $E_6$ grand unified theory with
  $SU(2)_{\sf H}$ family symmetry, the spontaneous $CP$ violation can
  solve the supersymmetric $CP$ problem.
  The scenario predicts $V_{ub} \sim \mathcal{O}(\lambda^4)$ rather
  than $\mathcal{O}(\lambda^3)$ which is naively expected value,
  because of a cancellation at the leading order.
  Since the experimental value of $V_{ub}$ is $\mathcal{O}(\lambda^4)$,
  it must be important to consider the reason and the conditions for
  the cancellation.
  In this paper, we give a simple reason for the cancellation and
  show that in some $E_6$ models such a cancellation requires that
  the vacuum expectation value (VEV) of the adjoint Higgs does not
  break $U(1)_{B-L}$.
  Note that this direction of the VEV plays an important role
  in solving the doublet-triplet splitting problem by
  Dimopoulos-Wilczek mechanism.
  In this $E_6$ models, the experiments may measure the direction of
  the adjoint Higgs VEV by measuring the size of
  $V_{ub} \sim \mathcal{O}(\lambda^4)$.
 \end{abstract}

\end{titlepage}
\section{Introduction}
Grand unified theory (GUT) has several attractive features
\cite{Georgi:1974sy,Fritzsch:1974nn}.
It unifies not only three forces but also matter (quark and lepton)
fields in the standard model (SM). 
As the result of the unification, the origin of the hypercharge,
$U(1)_Y$, in the SM can be understood in the GUT.
The unification of three gauge interactions has quantitatively been
tested by calculating three running gauge couplings in the 
(supersymmetric) SM which meet at a scale (the GUT scale).
The unification of matter fields roughly explains the various
hierarchical structures of quark and lepton masses and mixings by an
simple assumption that the $\bm{10}$ multiplets of $SU(5)$ induce the
hierarchical structures of Yukawa couplings.
Actually, in $SU(5)$ unification let us assume that the Yukawa couplings
which include the first two generation of $\bm{10}$ multiplets,
$\bm{10}_1$ and $\bm{10}_2$ have suppression factors,
$\epsilon_1$ and $\epsilon_2$
$(\epsilon_1 \ll \epsilon_2 \ll 1)$, respectively.
Then, we can understand not only that the mass hierarchy of up quarks
is the strongest, that of neutrinos is the weakest and those of the down
quarks and charged leptons are in between those of up quarks and
neutrinos, but also that the quark mixings are smaller than the neutrino
mixings.

One of the most attractive features of the $E_6$ unification
\cite{Gursey:1975ki,Achiman:1978vg,Barbieri:1980vc,Kugo:1994qr,
Irges:1998ax,Bando:1999km,Bando:2000gs} is that the
assumption for the Yukawa hierarchies can be derived in a reasonable
setup
\cite{Bando:2001bj,Maekawa:2002eh,Maekawa:2004qj}.
The derivation is so natural that even if we introduce a family
symmetry, $SU(2)_{\sf H}$ or $SU(3)_{\sf H}$, for  unifying the first
two generations or all three generations into a single multiplet,
$(\bm{27}, \bm{2})$ or $(\bm{27}, \bm{3})$, realistic Yukawa matrices
can be easily obtained by the spontaneous breaking of the family
symmetry \cite{Berezhiani:1985in,Maekawa:2002eh,Maekawa:2004qj}.
As the result of the unification of generations, SUSY flavor
problem \cite{Gabbiani:1996hi,Altmannshofer:2009ne} can be solved even
if large neutrino mixings are realized.

Recently, it has been pointed out that in the $E_6$ unification,
if $CP$ symmetry is spontaneously broken
\cite{Lee:1973iz,Maekawa:1992un,Pomarol:1992uu,Babu:1993qm}
by the vacuum expectation value which breaks the family symmetry
\cite{Barr:1988wk,Babu:1993ai,Nir:1996am,Ross:2004qn},
the Kobayashi-Maskawa (KM) phase \cite{Kobayashi:1973fv} can be
induced while the SUSY $CP$ problem is solved
\cite{Ishiduki:2009gr,Ishiduki:2009vr}.
Generically, in the non-universal sfermion masses which are predicted
by the family symmetry, complex Yukawa couplings, which are required
to produce non-vanishing KM phase, induce complex off-diagonal sfermion
masses which result in too large electric dipole moments (EDM) of the
neutron \cite{Hisano:2004tf}. 
It is non-trivial that this
problem can be solved by this spontaneous $CP$ violation. Moreover, it has
been pointed out that in the model \cite{Ishiduki:2009vr}, the $(1,3)$
component of the KM matrix, $V_{ub}$, becomes rather smaller value,
$\mathcal{O}(\lambda^4)$, than the expected value,
$\mathcal{O}(\lambda^3)$, because of a cancellation.
Here, $\lambda\sim 0.22$ is the Cabibbo angle \cite{Cabibbo:1963yz}.
Since the experimental value of $V_{ub}$ is of order $\lambda^4$ rather
than of order $\lambda^3$ \cite{Amsler:2008zzb}, it must be important to
study the reason of this cancellation and to know the conditions for
the cancellation in the $E_6$ GUT model.

In this paper, we examine the reason and the conditions for the
cancellation.
First, we show that in the $E_6$ GUT such cancellation is not special,
i.e., the quark mixings, not only $V_{ub}$ but also Cabibbo mixing,
tend to be small because of the unification of the Yukawa couplings.
Such cancellation due to Yukawa unification has been known in
$SO(10)$ GUT, but even in $E_6$ GUT, it has been discussed in the
literature \cite{Bando:2000gs}.
To avoid this cancellation, the mixing of doublet Higgses and/or the
higher dimensional interactions which include the $E_6$ Higgses
(adjoint Higgs and fundamental Higgs) are required.
It is not so easy that the effect of the higher dimensional interactions
becomes sizable.
One easy way is to introduce the anomalous $U(1)_A$ gauge symmetry
\cite{Witten:1984dg,Dine:1987xk,Atick:1987gy,Dine:1987gj},
with which the contribution of the higher dimensional
interactions becomes the same order as that of the tree level
interactions.
Another way is to introduce some symmetry which forbids the tree level
interactions and allows the higher dimensional interactions.
The family symmetry, $SU(2)_{\sf H}$, can play the role, if the first
two generation fields behave as doublet under the $SU(2)_{\sf H}$.
The $(1,2)$ component is produced not from the tree level interactions
but from the higher dimensional interactions which include the adjoint
$E_6$ Higgs. 
Second, we consider the effect of the direction of the $E_6$ adjoint VEV
in the $E_6$ model with the family symmetry.
It is shown that the cancellation requires that the adjoint VEV does not
break $U(1)_{B-L}$. Note that the direction of the VEV which remains
$U(1)_{B-L}$ plays an important role in solving the doublet-triplet
splitting problem by the Dimopoulos-Wilczek mechanism
\cite{Dimopoulos,Srednicki:1982aj,Barr:1997hq,Maekawa:2001uk,
Maekawa:2002bk,Maekawa:2003bb}. 

This paper is organized as follows.
In the next section, after a brief review for the $E_6$ model,
we investigate the reason for the cancellation.
In the section 3, we consider the effect of the direction of the 
adjoint Higgs VEV.
In section 4, we give a summary.
\section{Flavor structure of $E_6$ GUT models}
\label{sec:structure}
In this section, after a brief review of $E_6$ GUT scenario,
we show that in a simple setup in $E_6$ GUT, the Cabibbo mixing and 
$V_{ub}$ are vanishing at leading order.
And we discuss how to avoid the situation.

First, let us remind the basics of the $E_6$ group.
The fundamental representation has 27 components and the adjoint
representation has 78. 
These representations can be decomposed into the representation of
$SO(10) \otimes U(1)_{V^{\prime}}$ as
\begin{equation}
 \bm{27} = \bm{16}_1 + \bm{10}_{-2} + \bm{1}_4,\qquad
  \bm{78} = \bm{45}_0 + \bm{16}_{-3} + \bar{\bm{16}}_3 + \bm{1}_0
\end{equation}
where the subscript indicates the $U(1)_{V^{\prime}}$ charge of
each representation.
Similarly, $SO(10)$ representations can be decomposed in terms of
$SU(5) \otimes U(1)_V$ representations:
\begin{equation}
 \bm{16} = \bm{10}_{-1} + \bar{\bm{5}}_3 + \bm{1}_{-5},\qquad
  \bm{10} = \bm{5}_2 + \bar{\bm{5}}_{-2},\qquad
  \bm{45} = \bm{24}_0+\bm{10}_4+\bar{\bm{10}}_{-4}+\bm{1}_0.
\end{equation}
The subscript means the $U(1)_V$ charge in this case.

Since we must break $E_6$ gauge group into
the SM gauge group
$SU(3)_C \otimes SU(2)_L \otimes U(1)_Y \equiv G_{\text{SM}}$, 
we introduce two pairs of
(anti-)fundamental fields $\Phi$ ($\bar{\Phi}$), $C$ ($\bar{C}$) and
an adjoint Higgs field $A$.
Then we suppose that the vacuum expectation value (VEV) of $\Phi$
($\bar{\Phi}$) breaks $E_6$ into $SO(10)$ and $C$ ($\bar{C}$) breaks
$SO(10)$ into $SU(5)$:
\begin{equation}
 \braket{\Phi} = \braket{\bm{1}_{\Phi}} \neq 0,\qquad
  \braket{C} = \braket{\bm{16}_C[\ni \bm{1}_C]} \neq 0.
\end{equation}
In this paper, we often use a notation that the dimension of each
field is expressed by its name.
For example,
$\bm{27}_{\Phi} = \bm{16}_{\Phi} + \bm{10}_{\Phi} + \bm{1}_{\Phi}$ etc.
Here $|\braket{\Phi}| = |\braket{\bar{\Phi}}|$ and
$|\braket{C}| = |\braket{\bar{C}}|$ should be satisfied
from the $D$-flatness conditions.
On the other hand, since an adjoint Higgs $A$ must break
$SU(5)$ into $G_{\text{SM}}$, the VEV of the adjoint Higgs
$\braket{A}$ can be generally written as
\begin{equation}
 \label{eq:adj}
 \braket{A} = xQ_{V^{\prime}} + yQ_V + zQ_Y
\end{equation}
where $Q_{X}$ ($X = V^{\prime},V,Y$) stands for a generator corresponding
to the $U(1)_X$ charge. Note that $z \neq 0$ is needed to break $SU(5)$.

The matter fields in the standard model can be embedded in the
fundamental representations $\Psi_i$ with $i=1,2,3$.
Each generation of $\Psi_i$ contains $\bm{10}$ and $\bar{\bm{5}}$
representations of $SU(5)$:
\begin{equation}
 \Psi_i = \bm{16}_i[\bm{10}_i + \bar{\bm{5}}_i + \bm{1}_i]
  + \bm{10}_i[\bm{5}_i + \bar{\bm{5}}^{\prime}_i]
  + \bm{1}_i[\bm{1}^{\prime}_i]. \quad(i=1,2,3)
\end{equation}
To obtain the Yukawa couplings, we have to fix the origin of the SM
doublet Higgses.
Here, for economical reason, we assume that the minimal supersymmetric
SM (MSSM) Higgses come from $\Phi$ and $C$.
(It is natural that one of the MSSM Higgses come from $\Phi$ if we
consider how to realize doublet-triplet splitting naturally.
But here, we do not address about it.)
Then, we can obtain the Yukawa interactions from trilinear terms of
$\bm{27}$ fields:
\begin{equation}
 \label{eq:trilinear}
 W = (Y_{\Phi})_{ij}\Psi_i\Psi_j\Phi + (Y_C)_{ij}\Psi_i\Psi_jC.
\end{equation}
Note that there are six $\bar{\bm{5}}$ in three $\bm{27}$.
Three of six $\bar{\bm{5}}$ becomes superheavy with three ${\bm{5}}$
after breaking $E_6$ into $SU(5)$ through the above Yukawa interactions.
The mass term between $\bm{5}_i$ and $\bar{\bm{5}}_i$,
$\bar{\bm{5}}_i^{\prime}$ fields can be
expressed as $W_{\text{eff}} \supset \bm{5}M
(\bar{\bm{5}}^{\prime}, \bar{\bm{5}})^{\sf T}$ where
\begin{equation}
 \label{eq:mass}
  M \equiv (Y_{\Phi}, rY_C)\braket{\Phi},\qquad
  r \equiv \braket{C}/\braket{\Phi}.
\end{equation}
This means that the pairs of $\bm{5}$ and $\bar{\bm{5}}$ fields which
obtain GUT scale masses are decoupled at low energy, and the massless
modes are linear combinations of
$\bar{\bm{5}}_i$ and $\bar{\bm{5}}_i^{\prime}$.
If higher generation fields have larger Yukawa couplings, then two
$\bar{\bm{5}}$ fields from the third generation field ${\bm{27}_3}$
become superheavy unless $r\ll 1$. Therefore, it is reasonable that
all three SM $\bar{\bm{5}}$ fields come from the first two generation
fields, they have only smaller Yukawa couplings and milder hierarchical
structures for the Yukawa couplings of $\bar{\bm{5}}$ fields. 
This is an important point to produce realistic quark and lepton masses
and mixings in $E_6$ GUT.
The mixing of $\bar{\bm{5}}_i$ and $\bar{\bm{5}}^{\prime}_i$ for
these massless modes are determined from the diagonalizing procedure
\begin{equation}
 \label{eq:VMU}
 V^{\sf T}MU = \left(
  \begin{array}{@{\,}ccc|ccc@{\,}}
   m_1 &&&&& \\
   & m_2 &&& \bm{0}_{3 \times 3} & \\
   && m_3 &&&
  \end{array}
  \right)
\end{equation}
where $V$ and $U$ are unitary matrices which rotate $\bm{5}_i$ and
$(\bar{\bm{5}}_i, \bar{\bm{5}}^{\prime}_i)$ into their mass eigenstates,
respectively.
Provided that $m_i \neq 0$ ($i = 1,2,3$), there remains three massless
modes of $\bar{\bm{5}}$ fields and their mixing can be expressed as
\begin{equation}
 \label{eq:mixing}
 \begin{pmatrix}
  \bar{\bm{5}}^{\prime}_1 \\
  \bar{\bm{5}}^{\prime}_2 \\
  \bar{\bm{5}}^{\prime}_3 \\
 \end{pmatrix}
 = U_{\bm{10}}^0
 \begin{pmatrix}
  \bar{\bm{5}}^0_1 \\
  \bar{\bm{5}}^0_2 \\
  \bar{\bm{5}}^0_3 \\
 \end{pmatrix}
 + U_{\bm{10}}^{\sf H}
 \begin{pmatrix}
  \bar{\bm{5}}^{\sf H}_1 \\
  \bar{\bm{5}}^{\sf H}_2 \\
  \bar{\bm{5}}^{\sf H}_3 \\
 \end{pmatrix}
 ,\qquad
  \begin{pmatrix}
  \bar{\bm{5}}_1 \\
  \bar{\bm{5}}_2 \\
  \bar{\bm{5}}_3 \\
 \end{pmatrix}
 = U_{\bm{16}}^0
 \begin{pmatrix}
  \bar{\bm{5}}^0_1 \\
  \bar{\bm{5}}^0_2 \\
  \bar{\bm{5}}^0_3 \\
 \end{pmatrix}
 + U_{\bm{16}}^{\sf H}
 \begin{pmatrix}
  \bar{\bm{5}}^{\sf H}_1 \\
  \bar{\bm{5}}^{\sf H}_2 \\
  \bar{\bm{5}}^{\sf H}_3 \\
 \end{pmatrix}
\end{equation}
where $\bar{\bm{5}}^0_i$ and $\bar{\bm{5}}^{\sf H}_i$ represent
massless modes and heavy modes, respectively.
The $3 \times 3$ mixing matrices $U_{\bm{10}}^0$, $U_{\bm{10}}^{\sf H}$,
$U_{\bm{16}}^0$, $U_{\bm{16}}^{\sf H}$ are related with the
unitary matrix $U$ which appears in Eq. \eqref{eq:VMU} so that
\begin{equation}
 U \equiv \left(
  \begin{array}{@{\,}c|c@{\,}}
   U_{\bm{10}}^{\sf H} & U_{\bm{10}}^0 \\ \hline
    U_{\bm{16}}^{\sf H} & U_{\bm{16}}^0
  \end{array}
  \right).
\end{equation}
Note that by solving the following equation,
\begin{equation}
 \label{eq:MU}
 M\binom{U_{\bm{10}}^0}{U_{\bm{16}}^0}
  = (Y_{\Phi}U_{\bm{10}}^0 + rY_CU_{\bm{16}}^0)\braket{\Phi} = 0
\end{equation}
$U_{\bm{10}}^0$ and $U_{\bm{16}}^0$ can be determined
up to the multiplication of a $3\times 3$ unitary matrix from the right.
Because of the unitarity condition of $U$, three 6 dimensional vectors
$(u_i)_k$ $(i=1,2,3, k=1,2,\cdots,6)$, which are defined as 
$(u_i)_j = (U_{\bm{10}}^0)_{ij}, (u_i)_{j+3} = (U_{\bm{16}}^0)_{ij}$,
are normalized and orthogonal.
When the rank of $Y_\Phi$ equals to three,
it is obvious that three independent vectors satisfied with
Eq. \eqref{eq:MU} can be obtained by $(u_i)_{j+3} = \delta_{ij}$ and
$(u_i)_j = r(Y_\Phi^{-1}Y_C)_{ij}$.
Once we fix the vector space which satisfies Eq. \eqref{eq:MU},
we can determine the explicit normalized and orthogonal basis for the
space at any accuracy.
Then the MSSM Yukawa interaction terms
$(Y_u)_{ij}Q_i\bar{u}_jH_u$, $(Y_d)_{ij}Q_i\bar{d}_jH_d$ and
$(Y_e^{\sf T})_{ij}L_i\bar{e}_jH_d$ are obtained through
\begin{align}
 Y_u&: \bm{10}_i\bm{10}_j\bm{5}_{H_u}, \\
 Y_d, Y_e^{\sf T}&: \bm{10}_i\bar{\bm{5}}_j\bar{\bm{5}}^{\prime}_{H_d}
  + \bm{10}_i\bar{\bm{5}}_j^{\prime}\bar{\bm{5}}_{H_d}
  \to \bm{10}_i(U_{\bm{16}}^0\bar{\bm{5}}^0)_j\bar{\bm{5}}^{\prime}_{H_d}
  + \bm{10}_i(U_{\bm{10}}^0\bar{\bm{5}}^0)_j\bar{\bm{5}}_{H_d}.
 \label{eq:YdYe}
\end{align}

Now let us examine the feature of quark mixings in this model.
First, we consider a simple case where the MSSM Higgs fields
$H_u$ and $H_d$ are contained in ${\bm{10}_\Phi}$.
Then the Yukawa matrices for up- and down-quark can be written as
\begin{equation}
 Y_u = Y_{\Phi},\qquad
  Y_d = Y_{\Phi}U_{\bm{16}}^0.
\end{equation}
Without loss of generality, we can take the diagonal matrix $\hat{Y}_\Phi$
as $Y_\Phi$.
Let us consider a situation that 
the three massless modes of $\bar{\bm{5}}$ fields are mainly
composed of $\bar{\bm{5}}_1$, $\bar{\bm{5}}_1^{\prime}$ and
$\bar{\bm{5}}_2$.
As noted before, the two $\bar{\bm{5}}$ fields from the third generation
field tend to be become superheavy and therefore it is reasonable.
In such a case, by solving Eq. \eqref{eq:MU},  
$U_{\bm{10}}^0$ and $U_{\bm{16}}^0$ are approximately obtained as
\begin{equation}
 U_{\bm{10}}^0 \sim
  \begin{pmatrix}
   0 & 1 & 0 \\
   p_1 & p_2 & p_3 \\
   q_1 & q_2 & q_3
  \end{pmatrix}
  ,\qquad U_{\bm{16}}^0 \sim
  \begin{pmatrix}
   1 & 0 & 0 \\
   0 & 0 & 1 \\
   r_1 & r_2 & r_3
  \end{pmatrix}
\end{equation}
where $p_i, q_i, r_i \ll 1$ and their values are determined
by Eq. \eqref{eq:MU} (see Appendix B).
For example, when
\begin{equation}
\hat{Y}_{\Phi}\sim 
\begin{pmatrix}
   \lambda^6 & 0 & 0 \\
   0 & \lambda^4 & 0 \\
   0 & 0 & 1
\end{pmatrix}
,\qquad Y_C \sim
\begin{pmatrix}
   \lambda^6 & \lambda^5 & \lambda^3 \\
   \lambda^5 & \lambda^4 & \lambda^2 \\
   \lambda^3 & \lambda^2 & 1
\end{pmatrix}
\end{equation}
and $r \sim \lambda^{0.5}$, Eq. \eqref{eq:MU} becomes
\begin{equation}
U_{10}^0=
\begin{pmatrix}
   c_{11}\lambda^{0.5} & c_{21}\lambda^{-0.5} & c_{31}\lambda^{-2.5} \\
   c_{12}\lambda^{1.5} & c_{22}\lambda^{0.5} & c_{32}\lambda^{-1.5} \\
   c_{13}\lambda^{3.5} & c_{23}\lambda^{2.5} & c_{33}\lambda^{0.5}
\end{pmatrix}
U_{16}^0,
\end{equation}
where $c_{ij}$ are $\mathcal{O}(1)$ coefficients.
Taking account of the unitarity of $U$, $U_{10}^0$ and $U_{16}^0$
are calculated at the leading order as
 \begin{align}
 U_{\bm{10}}^0 &\sim
  \begin{pmatrix}
   \mathcal{O}(\lambda^{2.5}) & 1 & \mathcal{O}(\lambda^{1.5}) \\
   (c_{12} - \frac{c_{11}c_{32}}{c_{31}})\lambda^{1.5} &
   \frac{c_{32}}{c_{31}}\lambda &
   (c_{22} - \frac{c_{21}c_{32}}{c_{31}})\lambda^{0.5} \\
   (c_{13} - \frac{c_{11}c_{33}}{c_{31}})\lambda^{3.5} &
   \frac{c_{33}}{c_{31}}\lambda^3 &
   (c_{23} - \frac{c_{21}c_{33}}{c_{31}})\lambda^{2.5}
  \end{pmatrix}
  , \\
  U_{\bm{16}}^0 &\sim
  \begin{pmatrix}
   1 & 0 & 0 \\
   \mathcal{O}(\lambda^2) & 0 & 1 \\
   -\frac{c_{11}}{c_{31}}\lambda^3 & -\frac{1}{c_{31}}\lambda^{2.5} & 
   -\frac{c_{21}}{c_{31}}\lambda^2
  \end{pmatrix},
\end{align}
where three zero components are rotated away by the $3 \times 3$
unitary matrix from the right.
Then, we can obtain $Y_d$ as
\begin{equation}
 Y_d \sim
  \begin{pmatrix}
   \lambda^6 & 0 & 0 \\
   \mathcal{O}(\lambda^6) & 0 & \lambda^4 \\
    \lambda^3 & \lambda^{2.5} & \lambda^2
  \end{pmatrix}
\end{equation}
Since $Y_u = \hat{Y}_{\Phi}$ is diagonal, the $(1,3)$ and $(1,2)$
components of the CKM matrix are quite small in this setup.
Note that the situation does not change so much if the main part of
$\bar{\bm{5}}^0_3$ is $\bar{\bm{5}}^{\prime}_1$.
When the main part of $\bar{\bm{5}}^0_3$ is $\bar{\bm{5}}^{\prime}_1$,
all the CKM mixings become quite small.

There are basically two ways to avoid the smallness of the mixing angles
of the CKM matrix in $E_6$ models.
One way is to make the other Yukawa couplings like $Y_C$ contribute the
Yukawa couplings, $Y_u$ and $Y_d$ by, for example, introducing the
mixings of $\bm{27}_C$ into  the MSSM Higgs $H_u$ and/or $H_d$.
This is quite reasonable because we have $\bm{27}_C$ to break $SO(10)$
into $SU(5)$ in the $E_6$ GUT models, though it must be considered how
to obtain the mixing of $\bm{27}_C$ in solving the doublet-triplet
splitting problem.
The other way is to break the $E_6$ GUT relation for the Yukawa couplings
$Y_{\Phi}$ by, for example, introducing higher dimensional interactions
including $E_6$ Higgses which break $E_6$ into the SM gauge group. 
To avoid the undesirable $SU(5)$ GUT relation  $Y_d = Y_e^{\sf T}$,
we must introduce such higher dimensional interactions including
adjoint Higgs field $A$ such as
\begin{equation}
 W \supset \frac{1}{\Lambda}\Psi_i(A\Psi_j)H \qquad (H = \Phi, C)
\end{equation}
with a cutoff scale $\Lambda$.
Then after developing the VEV of $A$, the coefficients of the Yukawa
interactions do not respect $SU(5)$ symmetry though generically the
contributions from such higher dimensional interactions are suppressed
because the factor $\braket{A} / \Lambda$ is much smaller than one.
Generically, it is not easy that the contributions of such higher
dimensional interactions to the Yukawa couplings become sizable.
One of the easiest way to obtain sizable contributions of such higher
dimensional interactions is to introduce the anomalous $U(1)_A$ gauge
symmetry \cite{Maekawa:2001uk}.
Because the VEVs of fields $Z_i$ with $U(1)_A$ charges $z_i$ can be
obtained as
\begin{equation}
 \braket{Z_i} \sim
  \begin{cases}
   0 & (z_i > 0) \\
   (\xi / \Lambda)^{-z_i} & (z_i < 0)
  \end{cases}
  ,
\end{equation}
the higher dimensional interactions have the same order contributions
as the usual Yukawa interactions at tree level.
Here, $\xi$ is the parameter of the Fayet-Iliopoulos $D$-term.
Another way is to introduce family symmetry which forbids some of tree
Yukawa interactions and allows the higher dimensional interactions.
In next section, we discuss a concrete example in which such family
symmetry is introduced.
\section{CKM matrix in model with horizontal $SU(2)_{\sf H}$ symmetry}
\label{sec:horizontal}
\begin{table}
 \begin{center}
  \begin{tabular}{cccccccccc} \toprule
   & $\Psi_a$ & $\Psi_3$ & $F_a$ & $\bar{F}^a$ & $\Phi$ & $\bar{\Phi}$
   & $C$ & $\bar{C}$ & $A$ \\ \midrule
   $E_6 $ & $\bm{27}$ & $\bm{27}$ & $\bm{1}$ & $\bm{1}$ & $\bm{27}$
		       & $\bar{\bm{27}}$ & $\bm{27}$ & $\bar{\bm{27}}$
				   & $\bm{78}$ \\
   $SU(2)_{\sf H}$ & $\bm{2}$ & $\bm{1}$ & $\bm{2}$ & $\bar{\bm{2}}$
		   & $\bm{1}$ & $\bm{1}$ & $\bm{1}$ & $\bm{1}$
				   & $\bm{1}$ \\
   $Z_3$ & $0$ & $0$ & $1$ & $0$ & $0$ & $0$ & $2$ & $0$ & $0$
				       \\ \bottomrule
  \end{tabular}
 \end{center}
 \caption{Field contents and charge assignment under
 $E_6 \otimes SU(2)_{\sf H} \otimes Z_3$}
 \label{tab:fields}
\end{table}

In this section, we construct a simple model of flavor in $E_6$ GUT
and investigate the structure of CKM matrix in this model.
Here we focus on a model with a family symmetry $SU(2)_{\sf H}$
and spontaneous $CP$ violation.
Such a kind of model is examined in Ref. \cite{Ishiduki:2009vr} and we
consider a model which has the important features of the model.

First of all, we introduce a pair of $SU(2)_{\sf H}$ doublet fields
$F_a$ and $\bar{F}^a$ which is responsible for the breaking of the family
symmetry, $SU(2)_{\sf H}$.
Here the indices $a$ represent the transformation property in
$SU(2)_{\sf H}$ and it can be raised or lowered by means of the
antisymmetric symbols $\epsilon^{ab}$ and $\epsilon_{ab}$.
Taking into account the $D$-flatness condition and $SU(2)_{\sf H}$ gauge
degree of freedom, the VEV of $F_a$ and $\bar{F}^a$ can be generally
written as
\begin{equation}
 \braket{F} = \binom{0}{v_Fe^{i\delta}},\qquad
  \braket{\bar{F}} = \binom{0}{v_F}
\end{equation}
where we suppose that there appears nonzero phase $\delta$ and it breaks
the $CP$ invariance of the theory spontaneously.
As pointed out in Ref. \cite{Ishiduki:2009gr}, spontaneous $CP$
violation can be a natural solution of the SUSY $CP$ problem which may
arise in considering the effect of SUSY breaking.
That is, a real up-quark Yukawa matrix is favored for model building
because the non-universality of the up-squark mass with complex Yukawa
couplings leads to the problem of chromo-EDM constraint
\cite{Hisano:2004tf}.
To accomplish a real up-quark sector, we introduce a $Z_3$ discrete
symmetry and impose charge assignment shown in Table \ref{tab:fields}.
This $Z_3$ symmetry gives further advantage for the SUSY $CP$ problem,
because it prohibits dangerous terms of K\"{a}hler potential
(e.g., $K \supset \tilde{m}^2\theta^2\bar{\theta}^2\Psi_3^{\dagger}
(\Psi_aF^a)$) which induce complex squark masses.

With the above setup, we can write down the interaction terms which
contribute to the Yukawa matrices:
\begin{align}
 Y_{\Phi}: &
  \begin{pmatrix}
   0 & \Psi_a(A\Psi^a) & 0 \\
   \Psi_a(A\Psi^a) & (\Psi_a\bar{F}^a)^2 &
   (\Psi_a\bar{F}^a)\Psi_3 \\
   0 & (\Psi_a\bar{F}^a)\Psi_3 & \Psi_3\Psi_3
  \end{pmatrix}
  \Phi, \\
 Y_{C}: &
 \begin{pmatrix}
  0 & (\Psi_aF^a)(\Psi_b\bar{F}^b) & (\Psi_aF^a)\Psi_3 \\
  (\Psi_aF^a)(\Psi_b\bar{F}^b) & 0 & 0 \\
  (\Psi_aF^a)\Psi_3 & 0 & 0
 \end{pmatrix}
 C
\end{align}
where we adopt terms which give the leading contribution and
the higher dimensional terms are neglected.
In this paper, we often take the unit in which $\Lambda=1$.
Note that the terms like $\epsilon^{ab}\Psi_a\Psi_b\Phi$ do not exist
because $E_6$ singlet can be formed from the totally symmetric
combination of three $\bm{27}$s. 

In the following discussion, we assume that the MSSM Higgs fields
$H_u$ and $H_d$ are
\begin{equation}
 H_u \simeq \bm{10}_{\Phi}[\bm{5}_{\Phi}],\qquad
  H_d \simeq \bm{10}_{\Phi}[\bar{\bm{5}}^{\prime}_{\Phi}]
  + \alpha\bm{16}_C[\bar{\bm{5}}_C]
  + \beta\bm{10}_C[\bar{\bm{5}}^{\prime}_C]
\end{equation}
with $\alpha, \beta \ll 1$.
Here, there is no component from $C$ in $H_u$.
This is important in order to obtain real Yukawa matrix for up-type
quarks.
On the other hand, $\alpha\neq 0$ or $\beta\neq 0$ is required to
 obtain complex Yukawa
couplings for down-type quarks, which is important to obtain non
vanishing KM phase.
Now we write down the explicit form of the up-quark Yukawa matrix $Y_u$
as follows:
\begin{equation}
 \label{eq:Yu}
 Y_u = Y_{\Phi}^{(u)} \equiv
  \begin{pmatrix}
   0 & d & 0 \\
   -d & c & b \\
   0 & b & a
  \end{pmatrix}
  \sim
  \begin{pmatrix}
   0 & \lambda^A & 0 \\
   \lambda^A & \lambda^{2F} & \lambda^F \\
   0 & \lambda^F & 1
  \end{pmatrix}
  .
\end{equation}
where $a$, $b$, $c$, $d$ are real parameters.
Here the right hand side of Eq. \eqref{eq:Yu} indicates the order of each
element and we assume $\lambda^A \ll \lambda^{2F} \ll 1$.
Then $Y_u$ can be diagonalized by unitary transformations of
up-type quark fields as
$V_{uL}^{\sf T}Y_uV_{uR} = \diag(y_u,y_c,y_t)$.
The leading order contribution of $V_{uL}$ and $V_{uR}$ can be
calculated to be
\begin{equation}
 \label{eq:VuL}
 V_{uL} \simeq
  \begin{pmatrix}
   1 & \frac{ad}{ac - b^2} & 0 \\
   -\frac{ad}{ac - b^2} & 1 & \frac{b}{a} \\
   \frac{bd}{ac - b^2} & -\frac{b}{a} & 1
  \end{pmatrix}
  ,\qquad V_{uR} \simeq
  \begin{pmatrix}
   1 & -\frac{ad}{ac - b^2} & 0 \\
   \frac{ad}{ac - b^2} & 1 & \frac{b}{a} \\
   -\frac{bd}{ac - b^2} & -\frac{b}{a} & 1
  \end{pmatrix}
\end{equation}
and the Yukawa couplings $y_f \equiv m_f/v_u$ ($f = u,c,t$) are
obtained as
\begin{equation}
y_u \simeq \frac{ad^2}{ac - b^2},\qquad
y_c \simeq \frac{ac - b^2}{a},\qquad
y_t \simeq a.
\end{equation}
(See appendix \ref{sec:diagonalize} for this approximation.)

The structure of the down-quark sector is a little complicated because
of the mixing of $\bar{\bm{5}}$ fields shown in Eq. \eqref{eq:mixing}.
Since the contribution from massless modes $\bar{\bm{5}}_i^0$ can be
extracted in a manner shown in Eq. \eqref{eq:YdYe},
down-quark Yukawa matrix is
\begin{equation}
 \label{eq:Yd}
 Y_d = Y_{\Phi}^{(d)}U_{\bm{16}}^0 + \alpha Y_CU_{\bm{10}}^0
 + \beta Y_CU_{\bm{16}}^0
\end{equation}
where $Y_{\Phi}^{(d)} \neq Y_{\Phi}^{(u)}$ as mentioned in the previous
section.
That is, terms involving an adjoint
Higgs $A$ give different contribution for each component of the term:
\begin{equation}
 \Psi_a(\braket{A}\Psi^a)\Phi \ni (v_{\bar{u}} - v_Q)
  \epsilon^{ab}Q_a\bar{u}_bH_u + (v_{\bar{d}} - v_Q)
  \epsilon^{ab}Q_a\bar{d}_bH_d
\end{equation}
where $\braket{A}\psi \equiv v_{\psi}\psi$ for a component field $\psi$.
Thus, when $\Psi_a(A\Psi^a)\Phi$ gives a term $dQ_1\bar{u}_2H_u$,
there is a term $-(24\epsilon + 1)/5\cdot dQ_1\bar{d}_2H_d$ for
down-quark sector, where $\epsilon \equiv y/z$ with $y$ and $z$
are defined by Eq. \eqref{eq:adj}.
Since we have defined
$(Y_{\Phi}^{(u)})_{12} = -(Y_{\Phi}^{(u)})_{21} \equiv d$,
it leads to be $(Y_{\Phi}^{(d)})_{12} = -(Y_{\Phi}^{(d)})_{21} =
-(24\epsilon + 1)d/5$.
Moreover, the effect of an adjoint Higgs also appears on the mass
term between $\bm{5}$ and $\bar{\bm{5}}$ fields as
\begin{equation}
 \Psi_a(\braket{A}\Psi^a)\braket{\Phi} \ni (v_{\bar{d}^{\prime}} - v_d)
  \epsilon^{ab}d_a\bar{d}_b^{\prime}\braket{\bm{1}_{\Phi}}.
\end{equation}
Therefore we must use
$M_{d\bar{d}} \equiv (Y_{\Phi}^{(M_{d\bar{d}})}, rY_C)\braket{\Phi}$
with
$(Y_{\Phi}^{(M_{d\bar{d}})})_{12} = (Y_{\Phi}^{(M_{d\bar{d}})})_{21}
= (24\epsilon - 4)d/5$ instead of $M$ in Eq. \eqref{eq:mass}.
Taking them into account, one can write down the specific form of
the down-quark Yukawa matrix.
For this purpose, let us express the component of $Y_C$ as
\begin{equation}
 Y_C \equiv
  \begin{pmatrix}
   0 & f & e \\
   f & 0 & 0 \\
   e & 0 & 0
  \end{pmatrix}
  \sim
  \begin{pmatrix}
   0 & \lambda^{2F} & \lambda^F \\
   \lambda^{2F} & 0 & 0 \\
   \lambda^F & 0 & 0
  \end{pmatrix}
\end{equation}
where $\arg(e) = \arg(f) = \delta \neq 0$ and right-hand side of
equation represents the order of magnitude for each component.
If the main modes of $\bar{\bm{5}}^0_1$, $\bar{\bm{5}}^0_2$ and
$\bar{\bm{5}}^0_3$ are $\bar{\bm{5}}_1$, $\bar{\bm{5}}^{\prime}_1$
and $\bar{\bm{5}}_2$,
\footnote{
If the main modes $\bar{\bm{5}}^0_1$,
$\bar{\bm{5}}^0_2$ and $\bar{\bm{5}}^0_3$ are $\bar{\bm{5}}_1$,
$\bar{\bm{5}}_2$ and $\bar{\bm{5}}^{\prime}_1$, respectively,
then, the $V_{cb}$ is vanishing at leading order unless
$\alpha \sim \mathcal{O}(1)$.
If $\alpha \sim \mathcal{O}(1)$, generically, no cancellation happens.
}
respectively, then we can calculate $Y_d$ given by \eqref{eq:Yd}
to be
\begin{align}
 Y_d &\simeq
 \begin{pmatrix}
  - \alpha r\frac{af^2 + ce^2 - 2bef}{ac - b^2}
  - \beta\frac{24\epsilon - 4}{5}\frac{d(be - af)}{ac - b^2} &
  - \alpha\frac{24\epsilon - 4}{5}\frac{d(be - af)}{ac - b^2}
  - \beta\left(\frac{24\epsilon - 4}{5}\right)^2
  \frac{ad^2}{r(ac - b^2)} &
  - \frac{24\epsilon + 1}{5}d \\
  \frac{24\epsilon + 1}{5}d - \frac{24\epsilon - 4}{5}
  \frac{bd(be - af)}{e(ac - b^2)} + \beta f &
  - \left(\frac{24\epsilon - 4}{5}\right)^2\frac{abd^2}{re(ac - b^2)}
  + \alpha f &
  \frac{ce - bf}{e} \\
  - \frac{24\epsilon - 4}{5}\frac{ad(be - af)}{e(ac - b^2)}
  + \beta e &
  - \left(\frac{24\epsilon - 4}{5}\right)^2\frac{a^2d^2}{re(ac - b^2)}
  + \alpha e & \frac{be - af}{e}
 \end{pmatrix}
 \notag \\
 &\sim
 \begin{pmatrix}
  \alpha r\lambda^{2F} + \beta\lambda^A &
  \alpha\lambda^A + \beta r^{-1}\lambda^{2A - 2F} & \lambda^A \\
  \lambda^A + \beta\lambda^{2F} &
  r^{-1}\lambda^{2A - 2F} + \alpha\lambda^{2F} & \lambda^{2F} \\
  \lambda^{A - F} + \beta\lambda^F &
  r^{-1}\lambda^{2A - 3F} + \alpha\lambda^F & \lambda^F
 \end{pmatrix}
 .
\end{align}
Since $\lambda^A \ll \lambda^{2F} \ll 1$ is supposed, $Y_d$ has
a hierarchical structure $(Y_d)_{ij} \ll (Y_d)_{kj}$,
$(Y_d)_{ij} \ll (Y_d)_{il}$ with $i < k$, $j < l$ for a proper
choice of $\alpha$ and $r$.
So the unitary matrix which diagonalizes $Y_d$ as
$V_{dL}^{\sf T}Y_dV_{dR} = \diag(y_d,y_s,y_b)$ is
\begin{equation}
 \label{eq:VdL}
 V_{dL} \simeq
  \begin{pmatrix}
   1 & s_{3}^{dL*} & -\frac{24\epsilon + 1}{5}\frac{de}{be - af} \\
   -s_{3}^{dL} & 1 & \frac{ce - bf}{be - af} \\
   \frac{1}{be - af}
   \left[\frac{24\epsilon + 1}{5}de + s_3^{dL}(ce - bf)\right]
   & -\frac{ce - bf}{be - af} & 1
  \end{pmatrix}
\end{equation}
at leading order.
Here $s_{3}^{dL} \equiv \sin\theta_{3}^{dL}e^{i\chi_{3}^{dL}}$
is defined in Appendix A, and approximately given by
\begin{equation}
 s_{3}^{dL} \simeq
  \frac{(Y_d)_{12}(Y_d)_{33}-(Y_d)_{13}(Y_d)_{32}}
  {(Y_d)_{22}(Y_d)_{33}-(Y_d)_{23}(Y_d)_{32}}.
\end{equation}

Now we can calculate the CKM matrix element defined through
$V_{\text{CKM}} \equiv V_{uL}^{\dagger}V_{dL}$
using the explicit formulas Eqs. \eqref{eq:VuL} and \eqref{eq:VdL}:
\begin{equation}
 \label{eq:CKMleading}
 V_{\text{CKM}} \simeq
  \begin{pmatrix}
   1 & s_3^{dL*} - \frac{ad}{ac - b^2} & -\frac{24\epsilon + 6}{5}
   \frac{de}{be - af} \\
   -V_{us}^* & 1 & \frac{e(ac - b^2)}{a(be - af)} \\
   V_{ts}V_{cd} - V_{ub} & - V_{cb} & 1
  \end{pmatrix}
  \sim
  \begin{pmatrix}
   1 & \lambda^{A - 2F} & \lambda^{A - F} \\
   \lambda^{A - 2F} & 1 & \lambda^F \\
   \lambda^{A - F} & \lambda^F & 1
  \end{pmatrix}.
\end{equation}
To obtain a realistic flavor structure, we assume that
$\lambda^A \sim \lambda^5$, $\lambda^F \sim \lambda^2$,
$\alpha \sim \lambda^{0.5}$, $\beta\sim\lambda$ and $r \sim \lambda^{1.5}$,
which lead to
\begin{equation}
 Y_u \sim
  \begin{pmatrix}
   0 & \lambda^5 & 0 \\
   \lambda^5 & \lambda^4 & \lambda^2 \\
   0 & \lambda^2 & 1
  \end{pmatrix}
  ,\qquad Y_d \sim
  \begin{pmatrix}
   \lambda^6 & \lambda^{5.5} & \lambda^5 \\
   \lambda^5 & \lambda^{4.5} & \lambda^4 \\
   \lambda^3 & \lambda^{2.5} & \lambda^2
  \end{pmatrix}
  .
\end{equation}
If the VEV of the adjoint Higgs $\bm{45}_A$ is proportional to the
generator of $U(1)_{B - L}$, i.e., $\epsilon = - 1 / 4$,
the leading contribution to $V_{ub}$ vanishes.
(We discuss more general examples in which such cancellation happens
only when $U(1)_{B-L}$ is not broken by the adjoint Higgs VEV in
appendix \ref{sec:general}.)
The next-to-leading contribution to $V_{ub}$ can be estimated to be
\begin{equation}
 V_{ub} \simeq \frac{(Y_d)_{32}}{(Y_d)_{33}^2}
  \left[\frac{ad}{ac - b^2}(Y_d)_{22}e^{i\phi}
   + (Y_d)_{12}\right]
  \sim \mathcal{O}(\lambda^4),
\end{equation}
where $\phi = - 2\arg((Y_d)_{32}/(Y_d)_{33})$.
Therefore, we can obtain as
\begin{equation}
 V_{\text{CKM}} \sim
  \begin{pmatrix}
   1 & \lambda & \lambda^4 \\
   \lambda & 1 & \lambda^2 \\
   \lambda^3 & \lambda^2 & 1
  \end{pmatrix}.
  \label{eq:lambda}
\end{equation}
Since the measured value for $V_{ub}$ is $\mathcal{O}(\lambda^4)$
rather than $\mathcal{O}(\lambda^3)$, this result indicates that the
direction of the adjoint Higgs VEV is proportional to $U(1)_{B - L}$
in this $E_6$ GUT model.
On the other hand, this direction plays an important role in solving the
doublet-triplet splitting problem by Dimopoulos-Wilczek mechanism
\cite{Dimopoulos,Srednicki:1982aj}.
This coincidence is quite interesting and suggestive.
\section{Summary}
In this paper, we have examined how to obtain the CKM matrix
as in Eq. \eqref{eq:lambda} in an $E_6$ GUT with a family symmetry
$SU(2)_{\sf H}$.
Especially we have studied the reason for the cancellation by which
$V_{ub}$ becomes $\mathcal{O}(\lambda^4)$, not
$\mathcal{O}(\lambda^3)$.
Since the measured value of $V_{ub}$ is $\mathcal{O}(\lambda^4)$,
it must be valuable to know the reason for the cancellation.
First, we have shown that in a simple $E_6$ GUT model, 
the Cabibbo mixing and $V_{ub}$ tend to be much smaller
than the naively expected values because of the $E_6$ Yukawa unification.
Of course, in order to obtain sizable Cabibbo mixing, we must avoid such
suppression for the Cabibbo mixing. 
Therefore, we have studied several ways to avoid such suppression of
the CKM mixings.
Generically, when we introduce something to avoid the suppression,
it spoils the suppression of $V_{ub}$.
Actually, in the $E_6$ model with a family symmetry and spontaneous
$CP$ violation, generically, the suppression of $V_{ub}$ is spoiled.
Only when the adjoint VEV does not break $U(1)_{B-L}$, the cancellation
happens. 
Since such a direction of the VEV plays an important role in solving the 
doublet-triplet splitting problem by Dimopoulos-Wilczek mechanism,
it is interesting that the measured value of $V_{ub}$ is of order
$\lambda^4$.
It may be concluded that the experiments measured the direction of the
adjoint VEV in this $E_6$ GUT model with $SU(2)_{\sf H}$ by measuring
the size of $V_{ub}$. 
Since the direct search of the GUT is very difficult, indirect
measurements like this must be important.
Though one indirect measurement is not sufficient to confirm a 
scenario, we hope that various indirect measurements will confirm
some GUT scenario in future.
\section*{Acknowledgments}
N.M. is supported in part by Grants-in-Aid for Scientific Research from
MEXT of Japan.
This work was partially supported by the Grand-in-Aid for Nagoya
University Global COE Program,
``Quest for Fundamental Principles in the Universe:
from Particles to the Solar System and the Cosmos'',
from the MEXT of Japan.
\appendix
\section{Diagonalization procedure of the hierarchical matrix}
\label{sec:diagonalize}
In this appendix we summarize the procedure of diagonalization
of the $3 \times 3$ matrix $Y$ which is needed in computing the CKM
matrix.
Here we focus on the case where the matrix $Y$ has a hierarchical
structure $Y_{ij} \ll Y_{kj}$ and $Y_{ij} \ll Y_{il}$ with
$i < k$ and $j < l$. 
We mainly follow the procedure given in Ref. \cite{King:2002nf}.

The arbitrary matrix $Y$ can by diagonalized multiplying the unitary
matrices $V_L$ and $V^R$ as $V_L^{\sf T}YV_R = Y^{\sf D}$,
where $Y^{\sf D}$ is a diagonal matrix.
We can parametrize $V_L$ and $V_R$ so that
\begin{align}
 V_L^{\sf T} &\equiv
  \begin{pmatrix}
   c_3^L & -s_3^L & 0 \\
   s_3^{L*} & c_3^L & 0 \\
   0 & 0 & 1
  \end{pmatrix}
  \begin{pmatrix}
   c_2^L & 0 & -s_2^L \\
   0 & 1 & 0 \\
   s_2^{L*} & 0 & c_2^L
  \end{pmatrix}
  \begin{pmatrix}
   1 & 0 & 0 \\
   0 & c_1^L & -s_1^L \\
   0 & s_1^{L*} & c_1^L
  \end{pmatrix}
 \equiv P_3^LP_2^LP_1^L, \\
 V_R &\equiv
 \begin{pmatrix}
  1 & 0 & 0 \\
  0 & c_1^R & s_1^R \\
  0 & -s_1^{R*} & c_1^R
 \end{pmatrix}
 \begin{pmatrix}
  c_2^R & 0 & s_2^R \\
  0 & 1 & 0 \\
  -s_2^{R*} & 0 & c_2^R
 \end{pmatrix}
 \begin{pmatrix}
  c_3^R & s_3^R & 0 \\
  -s_3^{R*} & c_3^R & 0 \\
  0 & 0 & 1
 \end{pmatrix}
 \equiv P_1^{R\dagger}P_2^{R\dagger}P_3^{R\dagger}
\end{align}
where $s_i^{L/R} \equiv \sin\theta_i^{L/R}e^{i\chi_i^{L/R}}$ and
$c_i^{L/R} \equiv \cos\theta_i^{L/R}$ with $i = 1,2,3$.
If we write the components of $Y$ as
\begin{equation}
 Y \equiv
  \begin{pmatrix}
   y_{11} & y_{12} & y_{13} \\
   y_{21} & y_{22} & y_{23} \\
   y_{31} & y_{32} & y_{33}
  \end{pmatrix}
  ,
\end{equation}
the procedure of diagonalization is expressed as follows:
\begin{align}
 V_L^{\sf T}YV_R &= P_3^LP_2^LP_1^L
 \begin{pmatrix}
  y_{11} & y_{12} & y_{13} \\
  y_{21} & y_{22} & y_{23} \\
  y_{31} & y_{32} & y_{33}
 \end{pmatrix}
 P_1^{R\dagger}P_2^{R\dagger}P_3^{R\dagger}
 \simeq P_3^LP_2^L
 \begin{pmatrix}
  y_{11} & y^{\prime}_{12} & y_{13} \\
  y^{\prime}_{21} & y^{\prime}_{22} & 0 \\
  y_{31} & 0 & y_{33}
 \end{pmatrix}
 P_2^{R\dagger}P_3^{R\dagger} \notag \\
 &\simeq P_3^L
 \begin{pmatrix}
  y^{\prime}_{11} & y^{\prime}_{12} & 0 \\
  y^{\prime}_{21} & y^{\prime}_{22} & 0 \\
  0 & 0 & y_{33}
 \end{pmatrix}
 P_3^{R\dagger} =
 \begin{pmatrix}
  y^{\prime\prime}_{11} & 0 & 0 \\
  0 & y^{\prime}_{22} & 0 \\
  0 & 0 & y_{33}
 \end{pmatrix}
\end{align}
Here we supposed
$|s_i^{L/R}| \ll 1$ and $c_i^{L/R} \simeq 1$ and
\begin{equation}
 y^{\prime}_{22} \simeq y_{22} - \frac{y_{23}y_{32}}{y_{33}},\qquad
  y^{\prime}_{12} \simeq y_{12} - \frac{y_{13}y_{32}}{y_{33}},\qquad
  y^{\prime}_{21} \simeq y_{21} - \frac{y_{23}y_{31}}{y_{33}},
\end{equation}
\begin{equation}
 y^{\prime}_{11} \simeq y_{11} - \frac{y_{13}y_{31}}{y_{33}},\qquad
  y^{\prime\prime}_{11} \simeq y_{11}^{\prime}
  - \frac{y_{12}^{\prime}y_{21}^{\prime}}{y_{22}^{\prime}}
\end{equation}
which are good approximation if the hierarchy of the matrix $Y$ is
realized sufficiently.
Then the mixing angles for the left-handed field are
\begin{equation}
 \label{eq:sin}
 s_1^L \simeq \frac{y_{23}}{y_{33}},\qquad
  s_2^L \simeq \frac{y_{13}}{y_{33}},\qquad
  s_3^L \simeq
  \frac{y_{12}y_{33} - y_{13}y_{32}}{y_{22}y_{33} - y_{23}y_{32}}.
\end{equation}
Note that, if we consider the diagonalization of $2 \times 2$ matrix
\begin{equation}
 \begin{pmatrix}
  c_L & -s_L \\
  s_L^* & c_L
 \end{pmatrix}
 \begin{pmatrix}
  m_{11} & m_{12} \\
  m_{21} & m_{22}
 \end{pmatrix}
 \begin{pmatrix}
  c_R & s_R \\
  -s_R^* & c_R
 \end{pmatrix}
 =
 \begin{pmatrix}
  m^{\prime}_{11} & 0 \\
  0 & m^{\prime}_{22}
 \end{pmatrix}
\end{equation}
where $s_{L/R} \equiv \sin\theta_{L/R}e^{i\chi_{L/R}}$ and
$c_{L/R} \equiv \cos\theta_{L/R}$,
the explicit formula for $\tan 2\theta_{L/R}$ can be obtained as
\begin{align}
 \label{eq:tan}
 \tan 2\theta_L &=
 \frac{2(m_{12}m_{22} + m_{11}m_{21}e^{2i\chi_R})}
 {m_{22}^2e^{i\chi_L} - m_{11}^2e^{-i(\chi_L - 2\chi_R)}
 + m_{21}^2e^{i(\chi_L + 2\chi_R)} - m_{12}^2e^{-i\chi_L}}, \\
 \tan 2\theta_R &=
 \frac{2(m_{11}m_{12} + m_{21}m_{22}e^{2i\chi_L})}
 {m_{22}^2e^{-i(\chi_R - 2\chi_L)} - m_{11}^2e^{i\chi_R}
 - m_{21}^2e^{i(\chi_R + 2\chi_L)} + m_{12}^2e^{-i\chi_R}}.
\end{align}
Here $\chi_{L/R}$ are determined so that the right-hand side of
each equation becomes real.
Therefore the above approximation is correct if
$m_{22} \gg m_{11}, m_{12}, m_{21}$ in Eq. \eqref{eq:tan}.

We now write down the explicit expression for the CKM matrix elements
using the above parametrization.
Since the rotation matrix for left-handed up- and down-quarks are
\begin{equation}
 \label{eq:VL}
 V_{u/dL}^{\sf T} = P_3^{u/dL}P_2^{u/dL}P_1^{u/dL} \simeq
  \begin{pmatrix}
   1 & -s_3^{u/dL} & -s_2^{u/dL} + s_1^{u/dL}s_3^{u/dL} \\
   s_3^{u/dL*} & 1 & -s_1^{u/dL} \\
   s_2^{u/dL*} & s_1^{u/dL*} & 1
  \end{pmatrix}
\end{equation}
in this parametrization, CKM matrix element defined by
$V_{\text{CKM}} \equiv V_{uL}^{\dagger}V_{dL}$ can be obtained as
\begin{equation}
 \label{eq:VCKM}
 V_{\text{CKM}}^* \equiv V_{uL}^{\sf T}(V_{dL}^{\sf T})^{\dagger}
  \simeq
  \begin{pmatrix}
   1 & s_3^{dL} - s_3^{uL} & s_2^{dL} - s_2^{uL} - s_3^{uL}
   (s_1^{dL} - s_1^{uL}) \\
   -V_{us}^* & 1 & s_1^{dL} - s_1^{uL} \\
   V_{us}^*V_{cb}^* - V_{ub}^* & -V_{cb}^* & 1
  \end{pmatrix}
  .
\end{equation}
In Eqs. \eqref{eq:VL}, we supposed $s_2^{L/R} \ll s_1^{L/R}, s_3^{L/R}$
as can be expected from the experimental value for the CKM matrix
elements.
\section{Structure of CKM matrix in $E_6$ GUT}
\label{sec:general}
In this appendix, first, 
we give an explicit from of $U_{\bm{10}}^0$ and $U_{\bm{16}}^0$
when higher dimensional interactions are neglected.
Next, we show that the cancellation in $V_{ub}$ at the leading order
more generally happens when $U(1)_{B-L}$ is not broken by the adjoint VEV.

First we consider the superpotential given in Eq. \eqref{eq:trilinear}
neglecting the higher dimensional terms which involve the VEV of
some fields.
In such a case, we can always adopt the basis of the matter fields
$\Psi_i$ where $Y_{\Phi}$ is a diagonal matrix.
Then we express $Y_{\Phi}$ and $Y_C$ to be
\begin{equation}
 \label{eq:YPhiYC}
 Y_{\Phi} \equiv
  \begin{pmatrix}
   a_1 & & \\
   & a_2 & \\
   & & a_3
  \end{pmatrix}
  ,\qquad Y_C \equiv
  \begin{pmatrix}
   b_{11} & b_{12} & b_{13} \\
   b_{21} & b_{22} & b_{23} \\
   b_{31} & b_{32} & b_{33}
  \end{pmatrix}
\end{equation}
in this basis.
$U_{\bm{16}}^0$ is defined in Eq. \eqref{eq:mixing} and represents
the mixing of $\bar{\bm{5}}$ fields for the massless modes
$\bar{\bm{5}}_i^0$.
Then the mixing matrices $U_{\bm{10}}^0$ and $U_{\bm{16}}^0$ can be
calculated through Eq. \eqref{eq:MU} and expressed as
\begin{align}
 U_{\bm{10}}^0 &\simeq
  \begin{pmatrix}
   0 & 1 & 0 \\
   r(b_{23}b_{11} - b_{21}b_{13})/a_2b_{13} & a_1b_{23}/a_2b_{13}
   & r(b_{23}b_{12} - b_{22}b_{13})/a_2b_{13} \\
   r(b_{33}b_{11} - b_{31}b_{13})/a_3b_{13} & a_1b_{33}/a_3b_{13}
   & r(b_{33}b_{12} - b_{32}b_{13})/a_3b_{13}
  \end{pmatrix}
 , \\
 U_{\bm{16}}^0 &\simeq
 \begin{pmatrix}
  1 & 0 & 0 \\
  0 & 0 & 1 \\
  -b_{11}/b_{13} & -a_1/rb_{13} & -b_{12}/b_{13}
 \end{pmatrix}
\end{align}
Here we assume that the massless modes of $\bar{\bm{5}}$ fields
$\bar{\bm{5}}_1^0$, $\bar{\bm{5}}_2^0$, $\bar{\bm{5}}_3^0$ correspond
to $\bar{\bm{5}}_1$, $\bar{\bm{5}}_1^{\prime}$, $\bar{\bm{5}}_2$,
respectively, and
$(U_{\bm{10}}^0)_{2i}, (U_{\bm{10}}^0)_{3i}, (U_{\bm{16}}^0)_{3i} \ll 1$
with $i=1,2,3$.

Next we consider the more general situation where the higher dimensional
interaction terms also contribute to the Yukawa matrices.
As shown in Section \ref{sec:structure}, we must take into account the
fact that the VEV of an adjoint Higgs gives the different contribution
for the up-quark sector and down-quark sector.
Therefore we should express the Yukawa matrices as
\begin{equation}
 Y_u = Y_{\Phi}^{(u)},\qquad
  Y_d = Y_{\Phi}^{(d)}U_{\bm{16}}^0
\end{equation}
when 
$H_u \simeq \bm{10}_{\Phi}[\bm{5}_{\Phi}]$ and
$H_d \simeq \bm{10}_{\Phi}[\bar{\bm{5}}^{\prime}_{\Phi}]$.
The effective Yukawa matrices $Y_{\Phi}^{(u)}$ and $Y_{\Phi}^{(d)}$
are not equal in general.
Moreover, $Y_{\Phi}^{(u)}$ and $Y_{\Phi}^{(d)}$ are no longer guaranteed
to be symmetric matrices, and we cannot always make them diagonal
by the redefinition of the matter fields $\Psi_i$.
So the general expressions for $Y_{\Phi}^{(u)}$ and $Y_{\Phi}^{(d)}$ are
\begin{equation}
 Y_{\Phi}^{(u)} \equiv
  \begin{pmatrix}
   a^u_{11} & a^u_{12} & a^u_{13} \\
   a^u_{21} & a^u_{22} & a^u_{23} \\
   a^u_{31} & a^u_{32} & a^u_{33}
  \end{pmatrix}
  ,\qquad Y_{\Phi}^{(d)} \equiv
  \begin{pmatrix}
   a^d_{11} & a^d_{12} & a^d_{13} \\
   a^d_{21} & a^d_{22} & a^d_{23} \\
   a^d_{31} & a^d_{32} & a^d_{33}
  \end{pmatrix}
  .
\end{equation}
Since the mass terms between $\bm{5}$ and $\bar{\bm{5}}$ matter fields
also receive the different contribution from the higher dimensional
terms, $M \equiv (Y_{\Phi}^{(M)}, rY_C)\braket{\Phi}$ with
$Y_{\Phi}^{(M)}$ different from $Y_{\Phi}^{(u)}$ and $Y_{\Phi}^{(d)}$.

Although it is difficult to derive the expression of CKM matrix
in such a complicated case, we can show the characteristic feature
for some components of the CKM matrix.
To see this, we focus on the third column of the mixing matrix
$U_{\bm{16}}^0$:
\begin{equation}
 U_{\bm{16}}^0 \equiv
  \begin{pmatrix}
   \star & \star & x_1 \\
   \star & \star & x_2 \\
   \star & \star & x_3
  \end{pmatrix}
\end{equation}
where $x_1 \simeq 0$ and $x_2 \simeq 1$ if we assume that
$\bar{\bm{5}}_1^0$, $\bar{\bm{5}}_2^0$, $\bar{\bm{5}}_3^0$ are mainly
composed of $\bar{\bm{5}}_1$, $\bar{\bm{5}}_1^{\prime}$,
$\bar{\bm{5}}_2$ respectively.
Then the Yukawa matrix for the down-quark is
\begin{equation}
 Y_d = Y_{\Phi}^{(d)}U_{\bm{16}}^0 =
  \begin{pmatrix}
   \star & \star & a^d_{11}x_1 + a^d_{12}x_2 + a^d_{13}x_3 \\
   \star & \star & a^d_{21}x_1 + a^d_{22}x_2 + a^d_{23}x_3 \\
   \star & \star & a^d_{31}x_1 + a^d_{32}x_2 + a^d_{33}x_3
  \end{pmatrix}
  .
\end{equation}
From these information, we can see from Eq. \eqref{eq:sin} that
\begin{equation}
 \label{eq:sinuL}
 s_1^{uL} \simeq \frac{a^u_{23}}{a^u_{33}},\qquad
 s_2^{uL} \simeq \frac{a^u_{13}}{a^u_{33}},\qquad
 s_3^{uL} \simeq \frac{a^u_{12}a^u_{33} - a^u_{23}a^u_{13}}
 {a^u_{22}a^u_{33} - a^u_{23}a^u_{32}},
\end{equation}
\begin{equation}
 \label{eq:sindL}
 s_1^{dL} \simeq \frac{a^d_{21}x_1 + a^d_{22}x_2 + a^d_{23}x_3}
 {a^d_{31}x_1 + a^d_{32}x_2 + a^d_{33}x_3},\qquad
 s_2^{dL} \simeq \frac{a^d_{11}x_1 + a^d_{12}x_2 + a^d_{13}x_3}
 {a^d_{31}x_1 + a^d_{32}x_2 + a^d_{33}x_3},\qquad
\end{equation}
if $Y_u$ and $Y_d$ have a hierarchical structure which is supposed
in Appendix \ref{sec:diagonalize}.
Now we can obtain the specific expression for $V_{ub}$ from
Eq. \eqref{eq:VCKM}:
\begin{equation}
 \label{eq:Vub}
 V_{ub}^* \simeq s_2^{dL} - s_2^{uL} - s_3^{uL}
  (s_1^{dL} - s_1^{uL}).
\end{equation}
If we take $a^u_{ij} = a^d_{ij} \equiv a_{ij}$ in Eqs.
\eqref{eq:sinuL} and \eqref{eq:sindL}, Eq. \eqref{eq:Vub} can be
rewritten as
\begin{equation}
 V_{ub}^* \simeq
  \frac{x_1}{a_{31}x_1 + a_{32}x_2 + a_{33}x_3}
  \left[a_{11}
   + \frac{a_{12}(a_{23}a_{31} - a_{21}a_{33})
   + a_{13}(a_{21}a_{32} - a_{22}a_{31})}
   {a_{22}a_{33} - a_{23}a_{32}}\right]
\end{equation}
and $V_{ub} \propto x_1$ is derived.
Since $x_1 \simeq 0$ in our setup, the leading order contribution
for $V_{ub}$ vanishes if $a^u_{ij} = a^d_{ij}$ is satisfied.
On the other hand, if $a^u_{ij} \neq a^d_{ij}$, the non-vanishing
contribution to $V_{ub}$ generally appears.
For example, let us assume that $a^u_{12} \neq a^{d}_{12}$ and
$a^u_{21} \neq a^d_{21}$, which can be occurred by the presence of
the term $\Psi_a(A\Psi^a)\Phi$.
Here $\Psi_a$ is a horizontal $SU(2)_{\sf H}$ doublet matter field
which appears in Section \ref{sec:horizontal}.
Then, the additional contribution to $V_{ub}$ is
\begin{equation}
 \delta V_{ub}^* = -
  \frac{(a^u_{12} - a^d_{12})}{(a_{31}x_1 + a_{32}x_2 + a_{33}x_3)}x_2
\end{equation}
and it gives non-vanishing contribution because $x_2 \simeq 1$ if
the third generation of the massless $\bar{\bm{5}}$ field
$\bar{\bm{5}}_3^0$ is mainly $\bar{\bm{5}}_2$.
This is the case of Eq. \eqref{eq:CKMleading} and this contribution
vanishes if $a^u_{12} = a^d_{12}$ is satisfied.
Note that if the VEV of the adjoint Higgs field does not break
$U(1)_{B-L}$, then $a^u_{ij} = a^d_{ij}$ is satisfied and $V_{ub}$ is
vanishing at leading order. The size of $V_{ub}$ at sub-leading order
is dependent on the explicit model.
\end{document}